\documentclass[article,twocolumn,amsmath,floats,showpacs,nofootinbib]{revtex4}

\usepackage{graphicx}

\usepackage{textcomp}
\usepackage{hyperref}

\hypersetup{colorlinks=true}

\begin{document}

\title{Dependence of the amplitude of magnetoquantum oscillations of the metallic point-contact resistance on the bias voltage}

\author{N. L. Bobrov$^{1,2}$, N. N. Gribov$^{1,2}$, J. Kokkedee$^2$,   A. G. Jansen$^2$,  P. Wyder$^2$, I. K. Yanson$^1$}

\affiliation
{$^1$B.I.~Verkin Institute for Low Temperature Physics and Engineering, of the National Academy of Sciences of Ukraine, prospekt Lenina, 47, Kharkov 61103, Ukraine\\
$^2$High Magnetic Field Laboratory, Max-Planck Institut fur Festkorperforschung, and Centre National de la Recherche Scientifique, F-38042, Grenoble, France\\
Email address: bobrov@ilt.kharkov.ua}
\published {(\href{http://fntr.ilt.kharkov.ua/fnt/pdf/21/21-1/f21-0068r.pdf}{Fiz. Nizk. Temp.}, \textbf{21}, 68 (1995)); (Low Temp. Phys., \textbf{21}, 51 (1995)}
\date{\today}

\begin{abstract}Magnetoquantum oscillations of resistance as functions of the potential difference applied across the contact are studied for metallic point contacts made of $Al$ and $Be$. The amplitude of resistance oscillations in a magnetic field increases with voltage and is identical to the EPI spectrum of the point contact for one group of contacts, and with the bias voltage across the contact for another. The increase in the oscillation amplitude as well as its decrease has a nonmonotonic dependence on energy. The scattering of electrons by nonequilibrium phonons and the Fermi-liquid effects in the nonequilibrium electron system are considered as the possible reasons behind the observed effects.

\end{abstract}
\maketitle

\section{INTRODUCTION}
In a magnetic field \textbf{B}, the allowed electron states lie in the \textbf{k}-space on tubes called Landau tubes. These states are determined by the condition of quantization of the cross- section of a tube by a plane perpendicular to the direction of the magnetic field. If we take into consideration the tube with the largest cross-sectional area which partly lies inside the Fermi surface (FS), its filled part will decrease with increasing \textbf{B} and disappear at infinite velocity as the tube comes in contact with the FS. Such sharp decreases in the population density, which occur periodically in the reciprocal field $1/B$, lead to oscillations of the free energy and magnetization (de Haas-van Alphen effect dHvA). The frequency $F$ of these oscillations is described by the Lifshits-Onsager relation
\begin{equation} \label{eq__1}
{F=(c\hbar /2\pi e)A},
\end{equation}
where $A$ is the area of the extremal Fermi surface cross- section.

An increase in the temperature blurs the boundary between filled and unfilled states, and decreases the oscillation amplitude. The corresponding reducing factor has the form
\begin{equation} \label{eq__2}
{{R}_{T}}=\frac{2{{\pi }^{2}}{{k}_{B}}nT/\hbar {{\omega }_{c}}}{\text{sh}(2{{\pi }^{2}}{{k}_{B}}nT/\hbar {{\omega }_{c}})};
\end{equation}
where $n$ is the harmonic number in the oscillation spectrum, and ${{\omega }_{c}}=eB/m^{*}c$.

In addition, in view of the elastic scattering of electrons by impurities,\footnote{By impurities, we mean any static defects that scatter electrons elastically.} the momentum relaxation time becomes finite. According to the uncertainty principle, this leads to a broadening of the Landau levels and hence to the same extent of decrease in the oscillation amplitude as that caused by an increase in temperature from the actual value $T$ to ${{T}_{eff}}=T+x$. The corresponding reduction factor (the Dingle factor), which has the same meaning as the ratio of the actually observed amplitude of oscillations to the amplitude which would be observed in the absence of impurities, has the form
\begin{equation} \label{eq__3}
{{{R}_{\tau }}=\exp (-2{{\pi }^{2}}{{k}_{B}}nx/\hbar {{\omega }_{c}}),}
\end{equation}
where $x=\hbar /2{\pi}{k}_{B}\tau $ is the Dingle temperature, and  $\tau$ is the carrier lifetime.

It follows from the theory of noninteracting particles that all processes of electron scattering, including the electron- phonon collisions, should make a contribution to the Dingle temperature \cite{Shoenberg}. If higher-order contributions to the electron- phonon interaction (EPI) are disregarded, the effect of phonons boils down just to a decrease in the amplitude. The magnitude of the effect is determined in this case by the EPI function ${{\alpha}^{2}}(\omega)F(\omega)$  for the investigated metal, temperature, and magnitude of the magnetic field. However, a consideration of the many-particle effect shows that the scattering of electrons by phonons does not lead to an increase in the Dingle temperature \cite{Shoenberg}. Additional scattering of electrons by phonons upon an increase in temperature is compensated by a decrease in the mass ${{m}^{*}}={{m}_{0}}(1+\lambda)$  renormalized due to electron-phonon interaction ($\lambda $  is the EPI parameter), which makes this quantity nearly equal to the "bare" mass ${{m}_{0}}$. A departure from this formula towards higher amplitudes is observed in strong fields, and the required quantity \textbf{B} increases with temperature (phonon energy). It must be considered that, although EPI renormalizes the cyclotron mass determining the temperature attenuation of the amplitude, it does not affect the mass appearing in the Dingle factor, i.e., the correct expression \eqref{eq__3} contains the "bare" mass ${{m}_{0}}$ \cite{Shoenberg}.

The effect of EPI on the amplitude of dHvA oscillations can be studied in traditional experiments only for a limited number of metals (e.g., $Hg$) with an anomalously low Debye temperature, in which phonons can be excited at low temperatures when the oscillation amplitude is large enough for observation. Point contacts with a size of several tens or hundreds of angstroms provide a unique possibility of studying the effect of nonequilibrium phonons generated in a contact on the amplitude of oscillations of the electron density of states at the Fermi surface on the banks.

The simplest model of contact is a circular aperture in an infinitely thin opaque partition separating two metallic half-spaces. Owing to the small size $d$ of the point contact relative to the energy mean free path ${{l}_{\varepsilon }}$  of electrons, the electrons are divided into two groups whose Fermi energies differ by the applied potential difference $eV$ \cite{Kulik1, Kulik2}. In this case, the electrons from the group with a higher Fermi energy may go over to the group with a lower energy by emitting one phonon as a rule. Multiphonon processes are also possible, although their probability is much lower. Thus, if a voltage is applied across the contact, phonons with all possible energies right up to $eV$ will be generated in the contact region. Since ${{l}_{\varepsilon }}\gg d$, only a small fraction of total power is liberated in the immediate vicinity of the contact, and the temperature of the banks remains practically constant, viz., equal to the temperature of the helium bath.

In the prevailing technology for the creation of pressure point contacts, the elastic mean free path ${{l}_{i}}$, of electrons near the constriction is usually of the same order of magnitude as the contact diameter $d$, and much smaller than the mean free path at the banks. The contact resistance for zero bias voltage in the circular aperture model is defined by Wexler's interpolation formula \cite{Wexler}
\begin{equation} \label{eq__4}
{{{R}_{0}}=\frac{16\rho {{l}_{i}}}{3\pi {{d}^{2}}}+\beta ({{l}_{i}}/d)\frac{\rho }{d}}
\end{equation}
Here, ${\rho}l_i
={{p}_{F}}/{{e}^{2}}n=3/2N({{\varepsilon}_{F}}){{v}_{F}}{{e}^{2}}$  is constant for the given metal, ${{\varepsilon}_{F}},\text{ }{{p}_{F}},\text{ }{{v}_{F}}$ - Fermi energy, $\rho $  the resistivity, $N({{\varepsilon }_{F}})$  the electron density of states at the Fermi surface, and $\beta ({{l}_{i}}/d)$ is a nonmonotonic weakly varying function of ${{l}_{i}}/d$  whose value is close to unity: $\beta (0)=1$, $\beta \left( 4.48 \right)\simeq 0.6828$  (minimum of the function), $\beta (\infty )=9{{\pi }^{2}}/128\simeq 0.694$.

The first term in formula \eqref{eq__4} describes the Sharvin component of the point contact resistance, which does not depend on the electron mean free path and is determined by the shape of the Fermi surface of the given metal. The second term is the Maxwell component of resistance which depends on the purity of the metal near the constriction.

If the point contact is placed in a varying magnetic field, its resistance acquires an oscillatory component due to quantization of the electron energy spectrum. These oscillations may be associated with the Maxwellian or Sharvin component of resistance and depend on the parameters of the metal. In the contacts formed by conventional metals, the Farmor radii of electrons are much larger than the typical values of the contact diameter in fields up to $20\ T$. In the case of $Be$, for example, these quantities are comparable only for the largest contacts with $d\simeq 500\text{ }{\AA}$  and in the strongest field $10\ T$, i.e., for ${{r}_{B}}={{m}^{*}}{{v}_{F}}/eB=3000\ {\AA}$ (the value ${{m}^{*}}=0.17{{m}_{0}}$ was used in the estimates). Hence magnetoquantum oscillations of resistance of a metallic point contact are mainly associated with the fulfillment of the quantization conditions at the banks near the constriction. The oscillations of the electron density of states at the banks lead to the oscillations of the Sharvin component of the point-contact resistance. This is the main distinction between such oscillations and the oscillations of the resistance of point contacts formed by semimetals \cite{Gribov}, which are associated with the quantum oscillations of the collision integral \cite{Shoenberg} (oscillations of the Maxwellian component of resistance) and are analogous to the conventional Shubnikov-de Haas effect in bulk conductors.

\begin{figure}[]
\includegraphics[width=8cm,angle=0]{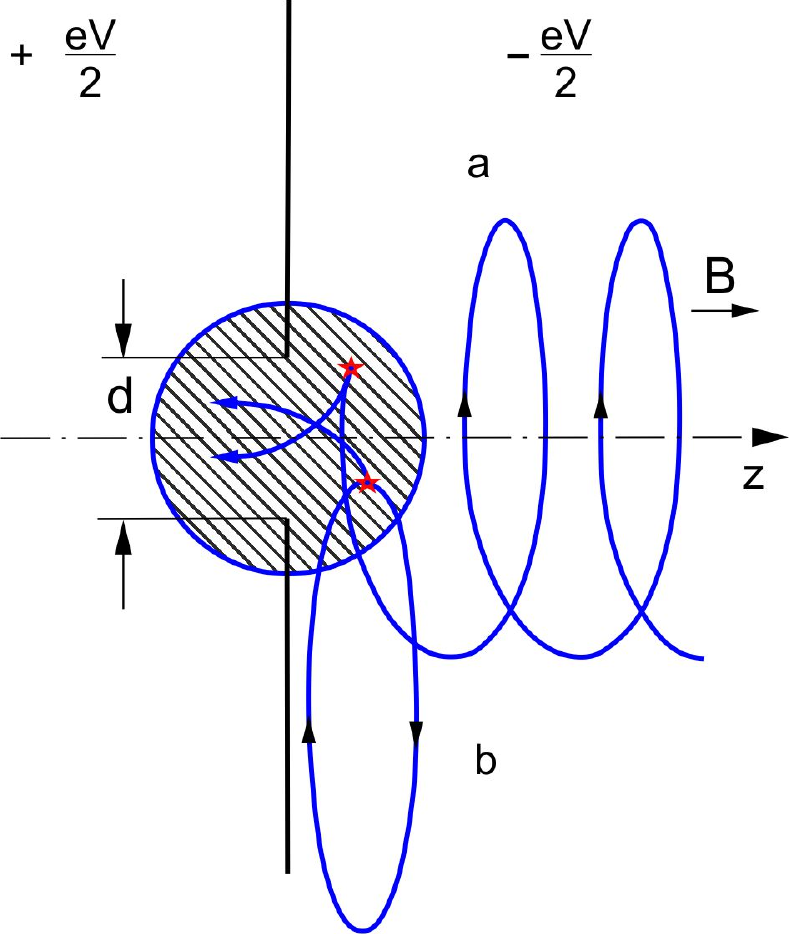}
\caption[]{Scattering of electrons moving in quantized orbits by nonequilibrium phonons (marked by stars) accumulated at the contact. The $z$-component of electron momentum and the ability to make an effective contribution to current increase after scattering. The case (a) corresponds to a small negative ${{v}_{z}}$ component and case (b) to zero ${{v}_{z}}$  component.}
\label{Fig1}
\end{figure}

The fact that the effect of impurities near the constriction is opposite to the effect of impurities at the banks of the contacts in the case of a magnetic field oriented parallel to the contact axis is paradoxical and leads to an \emph{increase} in the amplitude of the contact resistance oscillations \cite{Bogachek}. Indeed, the contribution to the magnetoquantum oscillations comes from the regions of extremal cross-section of the Fermi surface by the plane ${{p}_{z}}$=const (the $z$-axis is parallel to the contact axis and to the magnetic field direction). In ballistic contacts, such states correspond to low values of the component ${{v}_{z}}$  of the transport velocity of the carriers. Since the resistance of a point-contact is determined by the $z$-component of the transport velocity of electrons, the amplitude of oscillations of the Sharvin component of resistance is small. If the contact region contains elastic scatterers, the momenta of the electrons arriving in this region are distributed isotropically along the directions (Fig. \ref{Fig1}).
This results in an \emph{increase} in the $z$-component of the velocity of electrons from the extremal FS cross-section, and hence in an \emph{increase} in the amplitude of oscillations of the Sharvin component of the point-contact resistance. The presence of elastic scatterers at the contact banks leads to an increase in the Dingle temperature and hence to a \emph{decrease} in the oscillation amplitude.

The first experiments on the investigation of the effect of nonequilibrium phonons on the amplitude of the point- contact resistance oscillations were reported in Refs. \cite{Gribov} and \cite{Swartjes}. While describing the experimental results, the authors of Ref. \cite{Gribov} reduced the effect of phonons to overheating of the electron gas in the contact region, while two alternative mechanisms of damping were proposed in Ref. \cite{Swartjes}, viz., an increase in the Dingle temperature due to the electron- phonon scattering, or the Joule heating in the contact region.

In the present work, we study the effect of nonequilibrium phonons on the quantum magnetic oscillations in $Be$ and $Al$. For one group of contacts ($Al$ and $Be$), the amplitude of resistance oscillations in a magnetic field increases with voltage and is identical to the EPI spectrum of the point contact. For other point-contacts ($Al$), the magnetoresistance exhibits an increase in the oscillation amplitude with increasing potential difference across the contact. Both the increase in the oscillation amplitude and its decrease have a nonmonotonic energy dependence.
\section{EXPERIMENTAL TECHNIQUE AND PROCESSING OF RESULTS}
\begin{figure}[]
\includegraphics[width=8.7cm,angle=0]{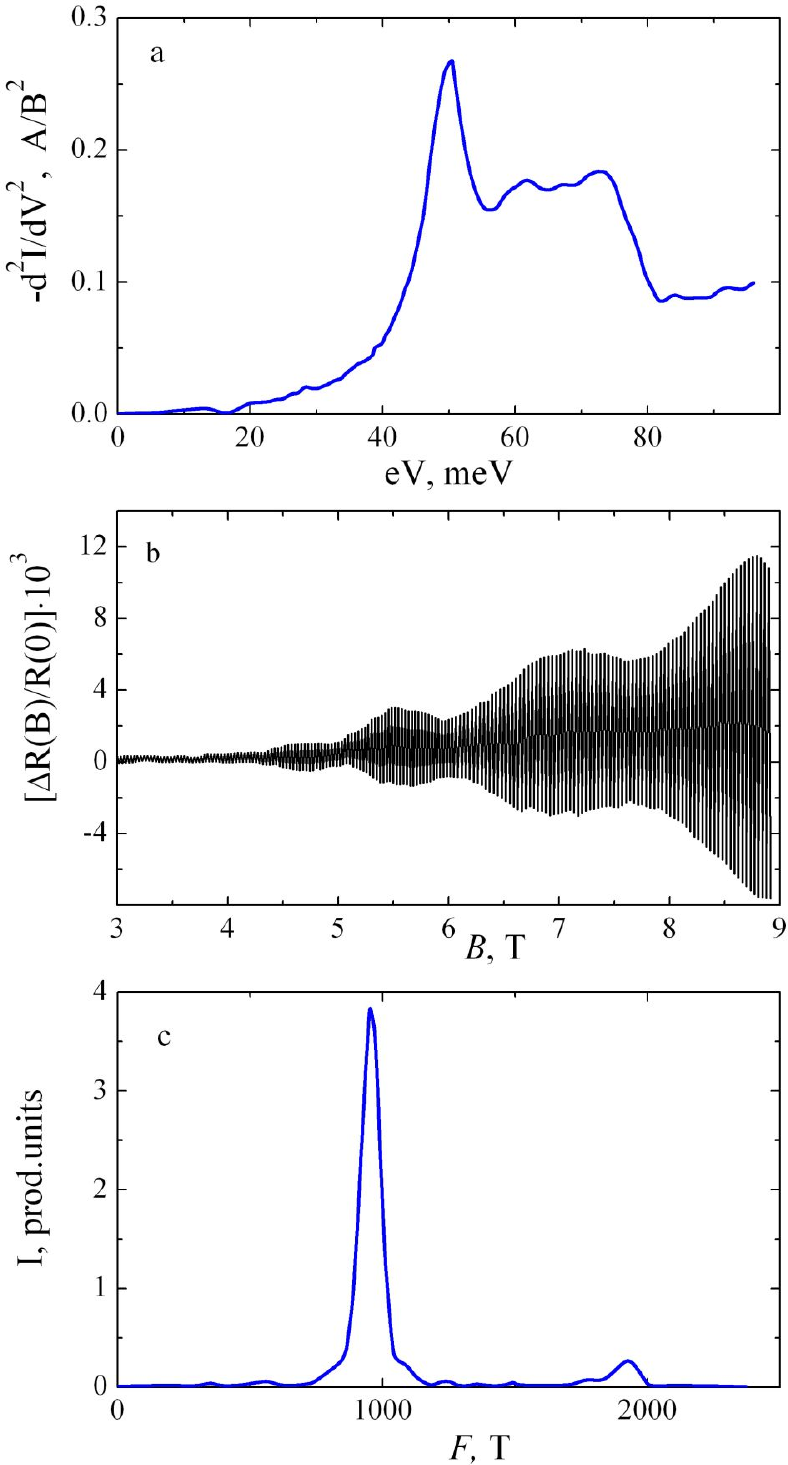}
\caption[]{(a) Point-contact spectrum of a $Be$ point contact, ${{R}_{0}}=16.24\text{ }\Omega $. \\(b) Oscillations of the resistance of a $Be$ point contact in a magnetic field at zero bias voltage $\mathbf{B}\parallel \mathbf{c}\parallel \mathbf{z}$, ${{R}_{0}}=16.24\text{ }\Omega $; $\Delta R/{{r}_{0}}\approx 2\cdot {{10}^{-2}}$ at $B\sim 9\text{ T}$. \\(c) $I$ and $F$ are the intensity and the frequency of oscillations.}
\label{Fig2}
\end{figure}

\begin{figure}[t]
\includegraphics[width=8.7cm,angle=0]{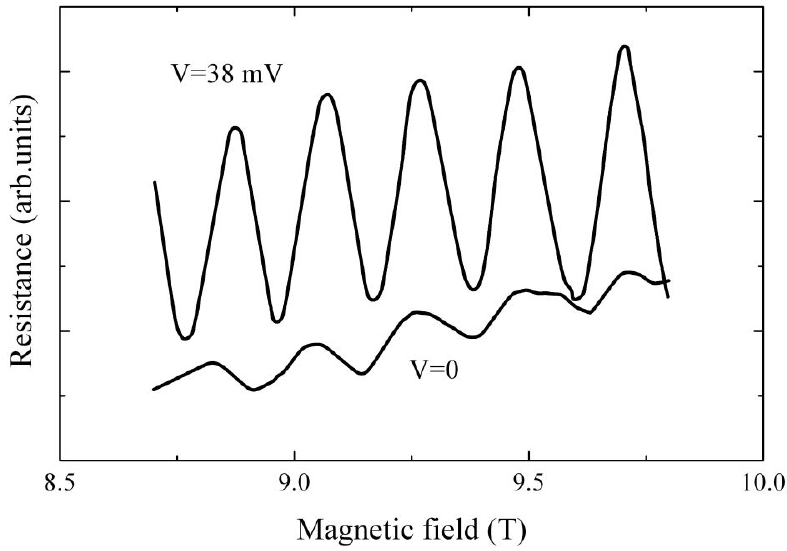}
\caption[]{Fragments of recording of magnetoresistance oscillations of an $Al$ point contact with $R_0=1.41\ \Omega$ for two different bias voltages.}
\label{Fig3}
\end{figure}

\begin{figure}[]
\includegraphics[width=8.7cm,angle=0]{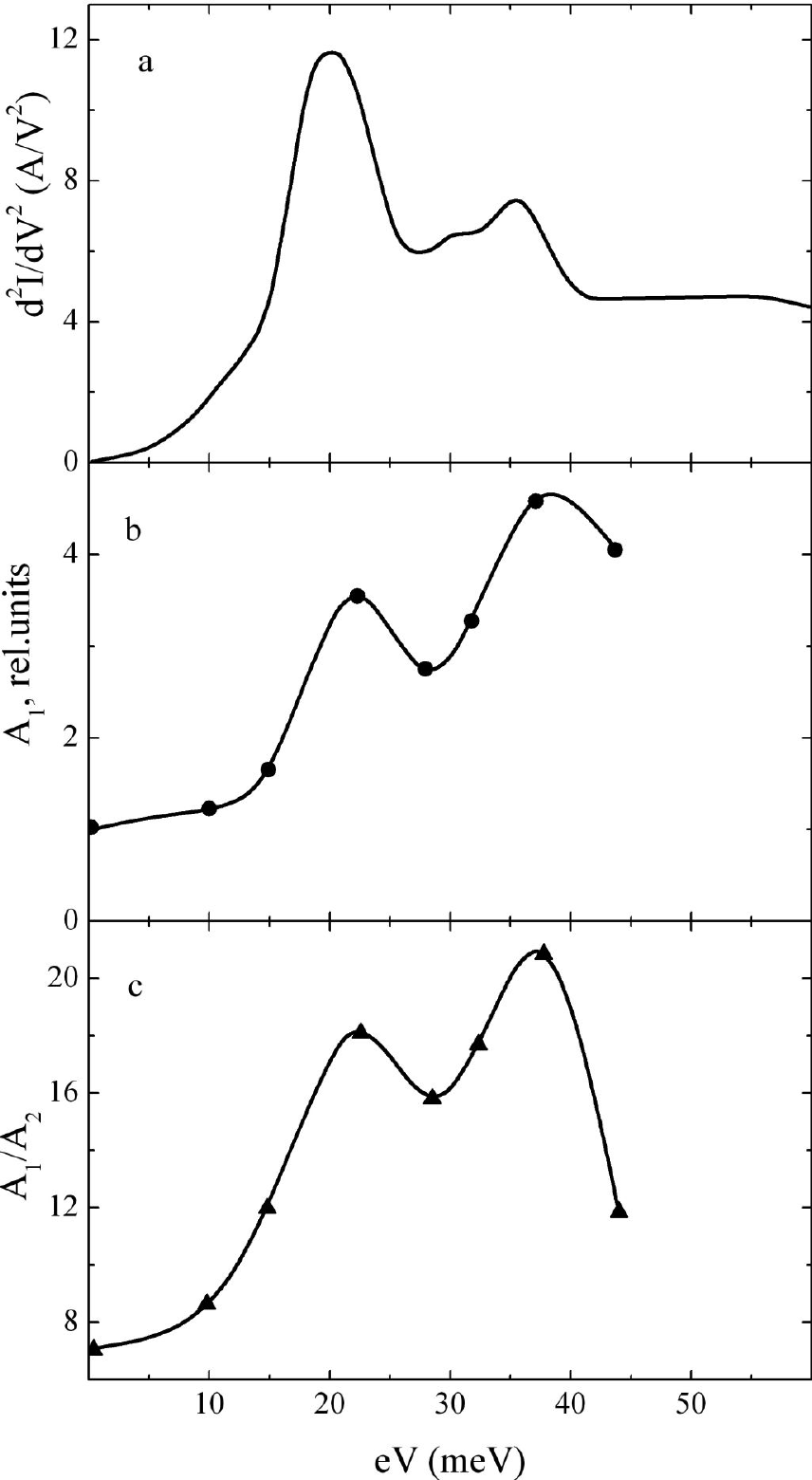}
\caption[]{\\(a) Point contact spectrum of an $Al$ point contact with $R_0=1.41\ \Omega $. \\(b) Relative change in the fundamental amplitude of resistance oscillations of a point contact in a magnetic field as a function of the applied bias voltage: $\Delta R/{{R}_{0}}\approx 5\cdot {{10}^{-4}}$  for $V=0$ and $B=9.5\ T$. \\(c) Dependence of the  amplitude ratio of the of the fundamental and the second harmonics as a function on $eV$.}
\label{Fig4}
\end{figure}

\begin{figure}[]
\includegraphics[width=8.7cm,angle=0]{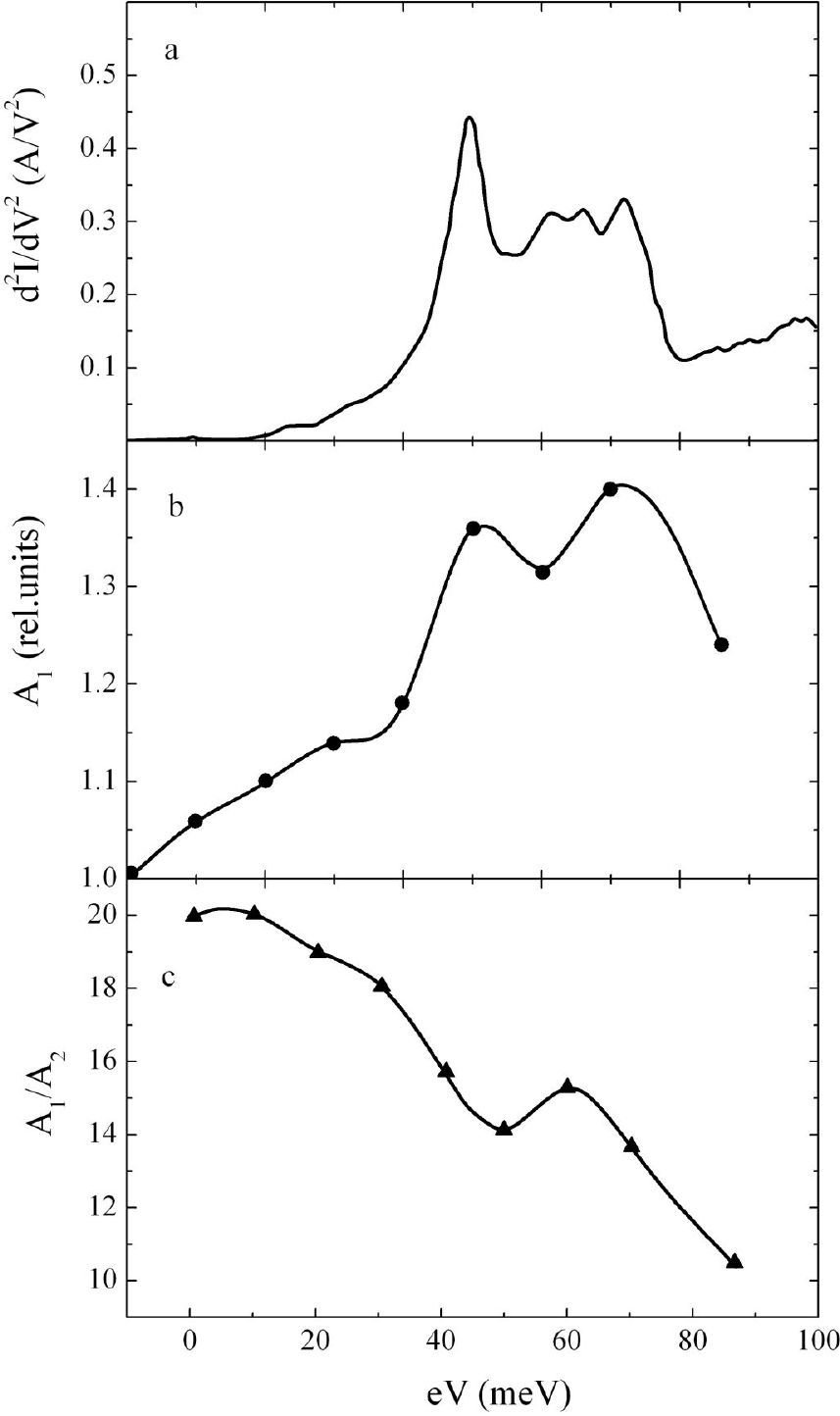}
\caption[]{\\(a) Point contact spectrum of an $Be$ point contact with $R_0=8.66\ \Omega $. \\(b) Relative change in the fundamental amplitude of resistance oscillations of a point contact in a magnetic field as a function of the applied bias voltage: $\Delta R/{{R}_{0}}\approx 2\cdot {{10}^{-3}}$  for $V=0$ and $B=9.5\ T$. \\(c) Dependence of the  amplitude ratio of the of the fundamental and the second harmonics as a function on $eV$.}
\label{Fig5}
\end{figure}

\begin{figure}[]
\includegraphics[width=8.7cm,angle=0]{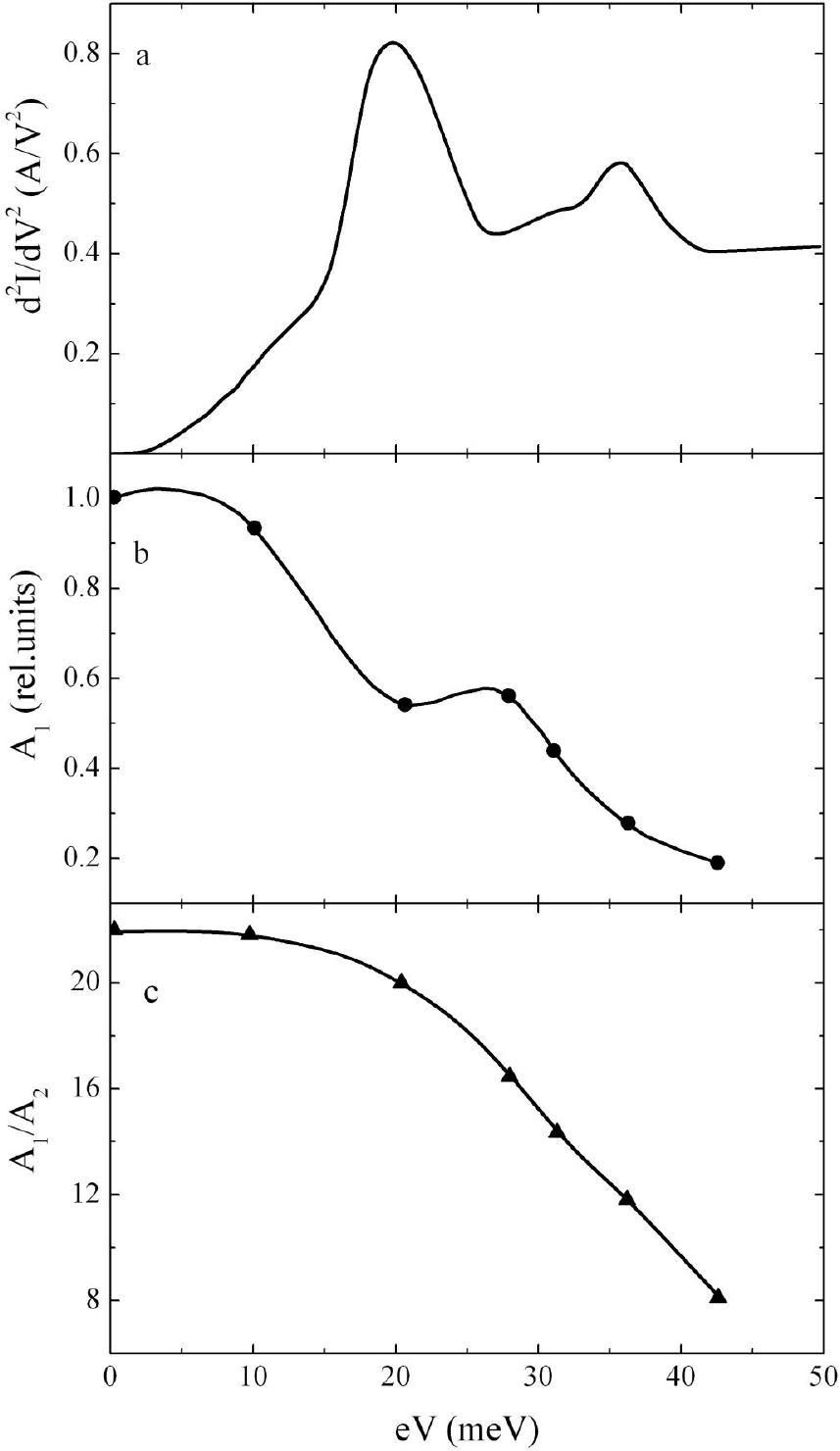}
\caption[]{\\(a) Point contact spectrum of an $Al$ point contact with $R_0=8.73\ \Omega $. \\(b) Relative change in the fundamental amplitude of resistance oscillations of a point contact in a magnetic field as a function of the applied bias voltage: $\Delta R/{{R}_{0}}\approx 3\cdot {{10}^{-3}}$  for $V=0$ and $B=9.5\ T$. \\(c) The ratio of the amplitudes of the fundamental and the second harmonics as a function of $eV$.}
\label{Fig6}
\end{figure}

Point-contacts were formed between the edges of two similarly oriented single-crystal electrodes of $Al$ or $Be$. These metals were chosen by us mainly because their energy spectrum contains groups of electrons with small effective masses, which makes it possible to observe the oscillations in relatively weak magnetic fields. In the case of $Be$, the contact axis coincides with the crystallographic axis $c$, while for $Al$ it is parallel to the (110) axis. The magnetic field was always applied parallel to the contact axis. The preliminary treatment of the electrodes included their chemical polishing in a mixture of acids for removing the defect layer formed during electric erosion cutting. For preparing $Al$ contacts, the electroforming \cite{Bobrov} and displacement techniques \cite{Chubov} were used to the same extent. The $Be$ contacts were formed predominantly by the electroforming technique which ensured their noticeably higher mechanical stability. We did not detect any correlation between the method of preparing a point contact and the form of the ${{A}_{1}}(eV)$  dependence. The first and second IVC derivatives were recorded by the standard modulation technique, and the data were registered directly on a computer. The following characteristics were recorded during measurements: ${{d}^{2}}V/d{{I}^{2}}(V);\ dV/dI(V);\ {{\left. dV/dI(B) \right|}_{V=const}}.$
The temperature was maintained at $1.3\ K$ throughout the measurements. By way of an example, Fig.\ref{Fig2}
 shows the PC spectrum of a $Be$ contact, the oscillating component of the magnetoresistance, and its Fourier spectrum. The PC spectrum ${{d}^{2}}I/d{{V}^{2}}\left( V \right)$ is obtained from the experimental formulas ${{d}^{2}}V/d{{I}^{2}}\left( V \right)$ and $dV/dI\left( V \right)$ by calculations and has a well-defined structure in the interval 40-80 $meV$, which is associated with the EPI function $\alpha _{pc}^{2}(\omega )F(\omega )$. The magnetoquantum oscillations of resistance with a natural frequency $985\ T$ correspond to the third electron band (cigar) and are known from the de Haas-van Alphen effect investigations \cite{Tripp}.
The Fourier spectrum also contains the second harmonic. The following procedure was used for determining the dependence of amplitude of magnetoresistance oscillations of the PC on bias voltage. For a fixed bias voltage across the contact, the oscillations ${{R}_{d}}(B)=dV/dI(B)$ were recorded in the same interval of magnetic fields ($8.5-10\ T$ as a rule). Figure \ref{Fig3}
 shows fragments of the recording for two different bias voltages applied to an $Al-Al$ point contact. For $Al$ contacts, the investigated oscillations correspond to the $\gamma$-pocket of the third band with a frequency $290\ \text{T}$ \cite{Larson}. Two recordings were made for each bias voltage, viz., in increasing and decreasing magnetic fields. After recording of the curves for the entire range of the values of $V$, a recording was made at $V=0$ and compared with the initial recording to ensure that the sample remained unchanged in the course of measurements. Contacts with different resistances (from 0.6 to 20\ $\Omega $), background levels (30 to 85\%) and levels of blurring of phonon singularities in the spectrum were studied. The complete set of characteristics was obtained for five $Al$ contacts and five $Be$ contacts. The results of experimental measurements were subjected to Fourier analysis for determining the amplitudes of oscillations in a fixed interval of magnetic fields for the entire set of curves. Since the characteristics $dV/dI(B)$ ($V$=const) were recorded for a constant value of the modulating current, and the differential resistance of the contact increases (up to 10\% in the interval of Debye energies) upon an increase in the bias voltage across the contact, the obtained results were reduced to a fixed modulating voltage and normalized to the value obtained for zero bias voltage across the contact. Figures \ref{Fig4}-\ref{Fig6}
  show the typical results of such calculations. For one group of point contacts, the amplitude $A_1$, of magnetoresistance oscillations increases with $eV$, and their energy dependence is identical to the EPI spectrum of the point contact (Figs. \ref{Fig4}, \ref{Fig5}). For another group of contacts ($Al$ contacts), the amplitude decreases monotonically as a function of voltage (Fig. \ref{Fig6}). The possible reasons behind such dependences will be discussed in the next section.
\section{DISCUSSION OF EXPERIMENTAL RESULTS}
Phonons are generated as a result of the relaxation of electrons injected from the contact by spontaneous emission processes. The resulting nonequilibrium distribution of phonons depends on the voltage applied to the contact. Like the effect of elastically scattered impurities, the scattering of electrons by nonequilibrium phonons affects the amplitude of quantum oscillations of the PC magnetoresistance in different ways depending on their position. The scattering of processes occurring in the region of the constriction lead to an increase in the amplitude, while scattering at the banks decreases its resultant value. We shall now discuss the dependence of the energy and coordinate distribution of phonons on the contact parameters and the effect of this distribution on magnetoquantum oscillations.

Apart from the defects that weakly affect the elastic mean free path of phonons but strongly scatter electrons, the spring-type contacts also contain, as a rule, a large number of line defects. These defects affect the momentum mean-free path $l_i$, of electrons weakly, but considerably shorten the
elastic mean free path $l_{i}^{ph}$  of nonequilibrium phonons generated by the electron flow through the aperture \cite{Kulik3}. If $l_{i}^{ph}<d$, the additional scattering of electrons by these nonequilibrium phonons accumulated in the contact leads to the emergence of a background (the energy-independent second IVC derivative ${{d}^{2}}I/d{{V}^{2}}\text{ }\left( eV \right)$  for $eV>\hbar {{\omega }_{D}}$ which is comparable with the spectral part in order of magnitude. A nonequilibrium phonon distribution is established in the contact with an effective temperature that depends on coordinates and phonon reabsorption coefficient. This distribution differs from the Planck distribution in that it contains a sharp edge at $\hbar \omega =eV$ \cite{Kulik3}. In this case, the nonequilibrium phonon distribution function is defined by the expression
\begin{equation} \label{eq__5}
{{{N}_{\omega }}(\mathbf{r})=\frac{eV-\hbar \omega }{2\hbar \omega }\Theta (eV-\hbar \omega ){{q}_{\omega }}(\mathbf{r}),}
\end{equation}
where
\begin{equation} \label{eq__6}
{{{q}_{\omega }}(\mathbf{r})\frac{1}{4\pi \Lambda _{\omega }^{2}}\int{d\mathbf{r}\frac{\exp (-R/{{\Lambda }_{\omega }})}{R}}q(\mathbf{{r}'});}
\end{equation}
\[\begin{matrix}
  R=\left| \mathbf{r}-\mathbf{{r}'} \right|; \\
  q(r)=\frac{1}{2}\left[ 1-{{\left( 1-\frac{\Omega (\textbf{r})}{2\pi } \right)}^{2}} \right]; \\
  {{\Lambda }_{\omega }}={{\left( l_{i}^{ph}l_{\varepsilon }^{ph}/3 \right)}^{1/2}}{{({{\omega }_{D}}-\omega )}^{{1}/{2}\;}} \\
\end{matrix}\]
Here $\Theta (x)$  is the Heaviside theta function; $l_{\varepsilon }^{ph}$  is the inelastic mean free path of phonons which has the same order of magnitude as the inelastic mean free path of electrons in the case of Debye energies and has the same energy dependence, $\Omega (\mathbf{r})$  is the solid angle at which the aperture is seen from the point \textbf{r}, and ${{\Lambda }_{\omega }}$  is the diffusive relaxation length over which a nonequilibrium phonon loses its energy.

\begin{figure}[]
\includegraphics[width=8.7cm,angle=0]{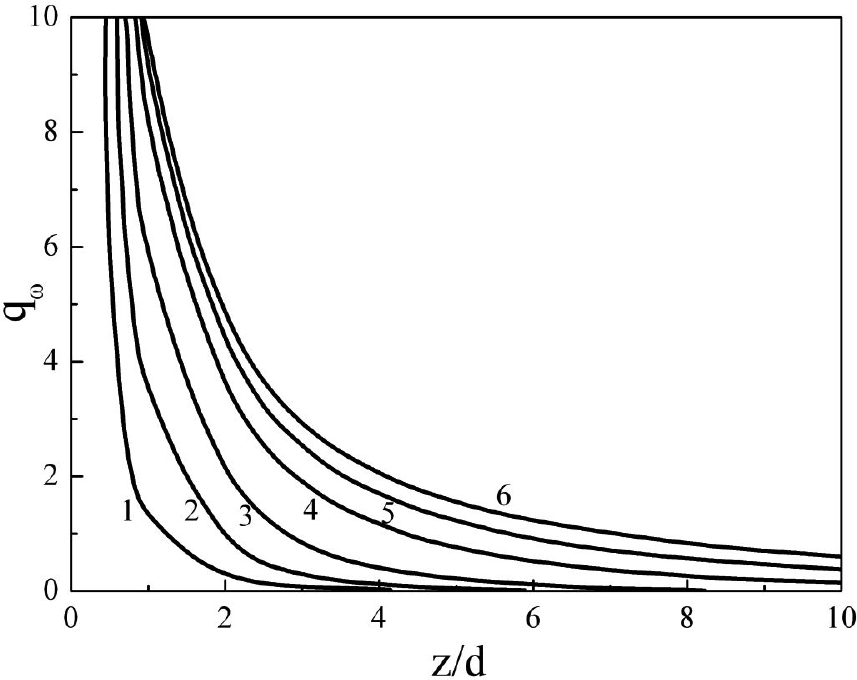}
\caption[]{Distribution of nonequilibrium phonons over the $z$-axis in the units of contact diameter as a function of their diffusive mean free path ${{\Lambda }_{\omega }}/d$, calculated by formula \eqref{eq__6}. All the curves corresponding to $z$=0 are normalized to unity; ${{\Lambda }_{\omega }}/d$=0.5 (curve 1); 1 (2); 2 (3); 5 (4); 10 (5); and 1000 (6).}
\label{Fig7}
\end{figure}

Thus, for a fixed value of $l_{i}^{ph}$, the diffusive relaxation length ${{\Lambda }_{\omega }}$ of phonons decreases with increasing phonon energy. Figure \ref{Fig7}
 shows a family of curves calculated from formula \eqref{eq__6} and describing the distribution of phonons along the $z$-axis (in the units of contact diameter) as a function of the parameter $\Lambda_\omega$. It can be seen that high-energy phonons having the smallest value of the ratio ${{\Lambda }_{\omega }}/d$ are accumulated near the aperture and are strongly reabsorbed by electrons while low-frequency phonons (with a large value of the ratio ${{\Lambda }_{\omega }}/d$) dominate at the banks.

All the contacts investigated by us lie in the spectral regime of current flow, i.e., satisfy the condition $d\ll {{\Lambda }_{\varepsilon }}$  where ${{\Lambda }_{\varepsilon }}={{\left( {{l}_{i}}{{l}_{\varepsilon }}/3 \right)}^{1/2}}$ is the diffusive length of energy relaxation of electrons (according to our estimates, $l_{\varepsilon D}^{Al}\sim 3000\ {\AA}$, $l_{\varepsilon D}^{Be}\sim 1500\ {\AA}$, ${{l}_{\varepsilon D}}={{v}_{F}}{{\tau }_{\varepsilon D}}$ ;\[\tau _{\varepsilon D}^{-1}=\frac{2\pi }{\hbar }\int\limits_{0}^{\omega D}{g(\omega )d\omega }\] at Debye energies \cite{Duif}. Hence a large part of the electrons loses its energy at the banks and not in the contact. In other words, the main source of nonequilibrium phonons at the banks is "hot" electrons injected from the contact. These phonons with randomly oriented momenta scatter electrons (including the electrons with extremal cross-sections of the Fermi surface) at the banks. The fraction of phonons arriving at the banks from the contact region is insignificant (see Fig. \ref{Fig7}).

Thus, we arrive at a complete analogy with impurities: scattering of electrons by phonons in the contact leads to an isotropization of the electron momenta. This leads to an increase in the $z$-component of the velocity of electrons from extremal FS cross sections and hence to an increase in the amplitude of oscillations of the contact resistance in a magnetic field. At the same time, the scattering of electrons by phonons generated by hot electrons at the banks lowers the amplitude of quantum oscillations of the density of states. The resulting dependence of the amplitude of point-contact resistance oscillations on $eV$ will be determined by the balance between these two processes.

The amplitude of quantum oscillations of the density of states at the banks depends nonlinearly on the concentration of the scatterers [see formula \eqref{eq__3}], attaining saturation for large values of  ${{\omega }_{c}}\tau $  and attenuating exponentially for small values. Hence if the concentration of static defects scattering electrons at the banks is low, the slight increase in the number of electron scattering acts by nonequilibrium phonons generated by "hot" electrons will lead to an insignificant decrease in the amplitude of quantum oscillations. However, if the concentration of static defects at the banks is quite high, a similar increase in the number of acts of electron scattering by nonequilibrium phonons may radically decrease the amplitude of oscillations at the banks. An increase in the contact size leads to an increase in the time of residence of nonequilibrium phonons near the constriction, and hence to an increase in the effectiveness with which quantized electrons are distributed isotropically along the directions as a result of phonon-electron collisions.

Hence we can expect an increase in the amplitude of oscillations with increasing bias voltage in low-resistance contacts with clean banks. Obviously, the strongest increase will be observed for contacts that are ballistic with respect to electrons and diffusive with respect to phonons, since they do not exhibit a background increase in the amplitude due to scattering of electrons by impurities, and a smaller elastic mean free path of nonequilibrium phonons is responsible for their higher concentration.

On the other hand, nonequilibrium phonons quickly leave the region of contacts that have a high resistance and are ballistic relative to phonons. If, on top of this, the contact banks are not very clean, a decrease in the amplitude of contact resistance oscillations will be observed with increasing $eV$ due to the influence of the Dingle factor at the contact banks.

These arguments have been confirmed in the experiments. For high-resistance $Al$ contacts (Fig. \ref{Fig6}) with a relatively low purity of banks near the constriction [this is confirmed by a large value of the ratio ${{A}_{1}}/{{A}_{2}}$  of harmonics for $V=0$, see formula \eqref{eq__8} below], the amplitude of resistance oscillations always decreases with increasing bias voltage across the contact. On the other hand, low-resistance $Al$ contacts with extremely clean banks (anomalously low value of the ratio ${{A}_{1}}/{{A}_{2}}$ of harmonics for $V=0$, see Fig. \ref{Fig4}), the dependence of the oscillation amplitude on $eV$ is similar to the PC spectrum. We obtained the dependences ${{R}_{d}}\left( H \right)$  for the given contact in the entire range of magnetic fields (0-10 T) for bias voltages of 0, 20, and 44 $mV$. Although the oscillations began for all three bias voltages in nearly the same magnetic fields, the increase in the oscillation amplitude with increasing bias voltage across the contact is more rapid (several times) in high magnetic fields (i.e., for large values of ${{\omega }_{c}}\tau $). A considerable (more than fourfold) increase in $A_1$ in strong fields and for large bias voltages, and its anomalously low value (about an order of magnitude smaller than the typical value) for $V=0$ also speaks in favor of a low concentration of impurities and static defects that scatter electrons directly in the contact. Irrespective of the variation of the resistance of $Be$ contacts with increasing bias voltage, we observed an increase in the amplitude of oscillations of $R_d$ (Fig. \ref{Fig5}). This is probably due to an extremely high rigidity of the $Be$ crystal lattice due to which the strains emerging during the contact formation do not move into the bulk of the material, but are rather concentrated near the surface and form effective reflectors which return the phonons to the contact. It is also necessary to use $Be$ single crystals with a high initial purity for preparing contacts.

The model presented above can explain not only the monotonic part of variation of the amplitude of oscillations with bias voltage, but also its similarity to the EPI spectrum. It was mentioned above that for $eV\le{{\omega }_{D}}$, the energy of the emitted phonons almost always coincides with the excess energy of electrons, due to a low probability of multiphonon processes. The group velocity of phonons generated by the electron flow depends on their energy and minimum for frequencies at which $\partial \omega /\partial q\sim 0$, i.e., near the peaks of the phonon density of states. Since ${{\Lambda }_{\omega }}$  depends on the velocity at which phonons leave the contact, a sharp increase in the phonon reabsorption coefficient will be observed at certain energies, as well as in the amplitude of the point-contact resistance oscillations. Since all electrons lying in the layer $eV$ can take part in phonon generation, the entire spectrum of phonons with energies ranging from 0 to $eV$ will be generated for a given bias voltage. Hence it can be expected that the dependence of the oscillation amplitude on $eV$ would be monotonic like the first IVC derivative $dV/dI\left( eV \right)$, which does not correspond to the experimental results. However, we have not taken into consideration the role of phonon- phonon collisions in this discussion.

The phonon mean free path $l_{ph}^{ph}$ defined by phonon- phonon collisions depends on the phonon energy and concentration which, in turn, is determined by the rate of phonon generation (i.e., by $eV$) as well as the rate at which they leave the contact. The latter depends on the group velocity of the phonons being generated, the contact diameter, and the phonon mean free paths ${{\Lambda }_{\omega }}$  and $l_{ph}^{ph}$ \cite{Kulik4}. Hence, for bias voltages that are higher than or of the same order as the phonon energies, the phonon-phonon collision frequency increases sharply due to a stepwise decrease in the diffusive energy relaxation length ${{\Lambda }_{\omega }}$ for phonons (similar to the energy relaxation length ${{\Lambda }_{\varepsilon}}$ for electrons), as well as to a sharp decrease in their group velocity. Such collisions lead to a thermalization of phonons, which may have three consequences.
\begin{enumerate}
  \item {The direction of the momentum of an electron may change significantly only as a result of its collision with a high-energy phonon. Hence thermalized phonons are less effective for a considerable increase in the $z$-component of the velocity of quantum electrons entering the contact.}
  \item {The phonon-phonon collisions lead to a decrease in the number of high-frequency phonons with low group velocities, which scatter the electrons most effectively, and the newly formed phonons will rapidly leave the contact region.}
  \item {An increase in the total number of phonons at the banks as a result of thermalization leads to a stronger suppression of oscillations in this region since collisions with phonons having any energy are effective in ejecting the quantized electrons from the extremal orbits.}
\end{enumerate}
Hence it can be expected that, for bias voltages that are higher than or of the same order as the characteristic phonon energies, the rise of the ${{A}_{1}}\left(eV\right)$ dependence will slow down and we may even obtain a descending region.

\begin{figure}[]
\includegraphics[width=8.7cm,angle=0]{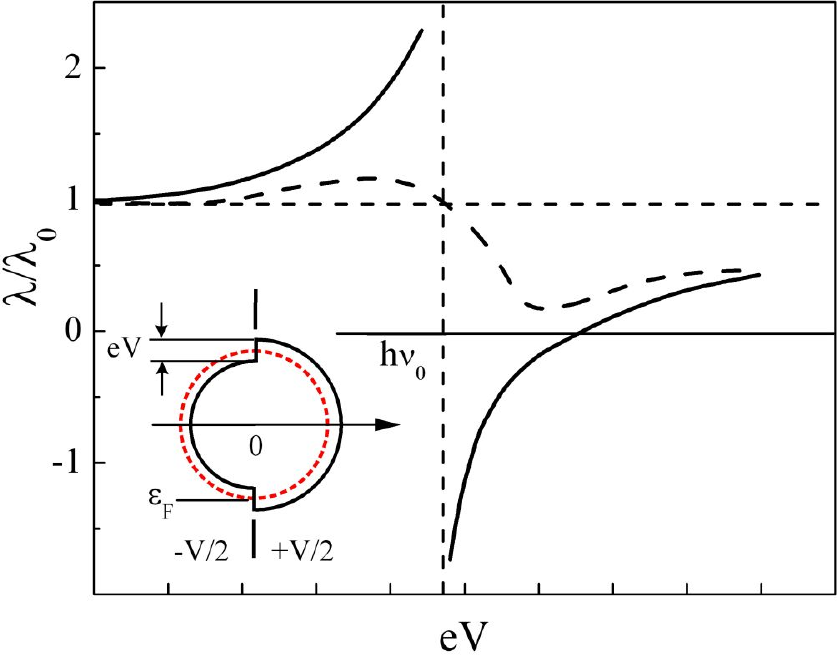}
\caption[]{Schematic representation of the behavior of the EPI parameter $\lambda$  near the aperture as a function of the applied voltage, which was predicted in Ref. \cite{Omel'yanchuk}. The solid and dashed curves correspond to an infinitely narrow phonon line and to a phonon band of a finite width with a characteristic energy $eV=h\nu_0$. The inset shows the electron distribution function for a ballistic point contact at the center of the aperture.}
\label{Fig8}
\end{figure}

Another possible factor contributing to the nonmonotonicity of ${{A}_{1}}\left(eV\right)$ may be the renormalization of the effective masses of high-energy electrons \cite{Omel'yanchuk} It was predicted in Ref. \cite{Omel'yanchuk} that, in the immediate vicinity of the contact, the EPI parameter $\lambda$  responsible for the renormalization of the effective mass ${{m}^{*}}=m(1+\lambda)$  differs from its equilibrium value ${{\lambda }_{0}}$. This difference is due to the presence of the nonequilibrium function of electron distribution in the point contact for $eV=0$. For a ballistic contact, $\lambda $ will depend on the applied voltage. For the Einstein model of spectrum of distribution of phonons with energy $h\nu_0$, the following expression holds at the center of the contact \cite{Omel'yanchuk}:
\begin{equation} \label{eq__7}
{\lambda (eV)=\frac{{{\lambda }_{0}}}{2}\left[ 1+\frac{{{(h{{\nu}_{0}})}^{2}}}{{{(h{{\nu}_{0}})}^{2}}-{{(eV)}^{2}}} \right].}
\end{equation}
The anticipated behavior of $\lambda (eV)$  is shown schematically in Fig. \ref{Fig8}.
It can be seen that for high voltages across the contact, the renormalization factor decreases by half since only half the Fermi surface participates in the renormalization of mass in the case of a nonequilibrium electron distribution function in the point contact (see inset to Fig. \ref{Fig8}). For $eV=h\nu_0$, resonance is observed in mass renormalization. The dashed curve in this figure shows the predicted form of the dependence $\lambda (eV)$ for the case of a blurred phonon line (band) near $h\nu$. It can be concluded from the theory \cite{Omel'yanchuk} that the nonmonotonic energy dependence of the effective electron mass for characteristic phonon frequencies is determined by the nonequilibrium electron distribution function in the PC for finite bias voltages.

The energy dependence of the renormalized effective mass in a ballistic point contact might be responsible for the observed nonmonotonic dependence of the amplitude of magnetoquantum oscillations on the applied voltage. It can be seen from Fig. \ref{Fig8} that the characteristic phonon energy should correspond to the point of inflection on the segment with a negative slope of the curve ${{A}_{1}}(eV)$. Although the exact position of the peak on the voltage dependence of the oscillations varies from contact to contact, the voltages corresponding to the points of inflection are usually higher than the values at which the phonon peaks are localized in the EPI spectrum of a point contact. As regards the observed disparity between the experimental and theoretical results, we note that the theory described in Ref. \cite{Omel'yanchuk} was developed for zero magnetic field. It would be interesting to include the energy dependence of the effective mass in the theory explaining the magnetoquantum oscillations observed in ballistic point contacts. The ballistic electron transport in a point contact makes it possible to study the energy dependence ${{m}^{*}}\left( eV \right)$  in the absence of thermal heating.

Let us now consider the dependence of the ratio ${{A}_{1}}/{{A}_{2}}$  of harmonic amplitudes on the bias voltage applied across the contact. For a fixed temperature, the ratio of amplitudes of the first and second harmonics is defined by the elastic mean free path ${{l}_{i}}={{v}_{F}}\tau $  of electrons in the region where the electron density of states is quantized in the metal, i.e., at the banks adjoining the contact. It follows from formulas \eqref{eq__2} and \eqref{eq__3} that
\begin{equation} \label{eq__8}
{{{A}_{1}}/{{A}_{2}}=C\cosh \left( \frac{2{{\pi }^{2}}{{k}_{B}}T}{\hbar {{\omega }_{c}}} \right)\exp \left( \frac{\pi }{{{\omega }_{c}}\tau } \right).}
\end{equation}
The constant $C$ has the meaning of the ratio of amplitudes of harmonics at $T=0$ in the absence of scatterers. Thus, as the electron mean free path decreases in the region of oscillations, the ratio ${{A}_{1}}/{{A}_{2}}$  increases. In view of the geometry of the experiments (the magnetic field is oriented parallel to the contact axis, while the electrons whose number oscillates have a small $z$-component of velocity) the ratio of the harmonics for $V=0$ contains information about the electron mean free path in the peripheral regions of the contact that are not far from the center of the contact, i.e., the regions where the electron mean free path is long enough for the emergence of oscillations. We assume that the form of the dependence of the ratio ${{A}_{1}}/{{A}_{2}}$ of the harmonics on the bias voltage applied across the contact is mainly due to two mechanisms:
\begin{enumerate}
  \item {A decrease in the (quasi)elastic mean free path of electrons in the region of space carrying information on the oscillations at $V=0$ due to an additional scattering by nonequilibrium phonons.}
  \item {A redistribution of the relative contribution to the information on electron mean free paths from spatial regions with different impurity concentrations.}
\end{enumerate}

Of course, both mechanisms can coexist. It is obvious that the first mechanism can lead only to an increase in the relative intensity of the first harmonic [see formula \eqref{eq__8}]. The second mechanism may lead to the opposite effect. Indeed, it was mentioned above that the amplitude of oscillations depends nonlinearly on the impurity concentration [formula \eqref{eq__3}]. We assume that regions with a low and a relatively high concentration of impurities exist in the vicinity of the point contact. In the current state, both types of regions will be exposed to phonons generated at the banks. Such a scattering only decreases the amplitude of density of states oscillations insignificantly in the case of clean regions, while the oscillations may be suppressed considerably in the dirty regions. Hence the relative contribution to the resistance oscillations from cleaner regions in space will increase with $eV$, and the ratio ${{A}_{1}}/{{A}_{2}}$ will decrease accordingly. Consequently, the shape of the dependence of ${{A}_{1}}/{{A}_{2}}$ on $eV$ can provide information about the distribution of impurities near the contact in the region where oscillations are observed. If the impurity concentration is equally low in all the regions where oscillations take place, the shape of the dependence of ${{A}_{1}}/{{A}_{2}}$ on $eV$ will on the whole be similar to the EPI spectrum (Fig. \ref{Fig4}). If, however, the impurities are distributed nonuniformly and their concentration in the oscillation region decreases with distance from the center of the contact, or if regions with a low impurity concentration and regions practically free from impurities exist at the same distance on the contact periphery, we shall observe a decrease in the ratio ${{A}_{1}}/{{A}_{2}}$ with increasing $eV$ (Figs. \ref{Fig5},\ref{Fig6}).
\section{CONCLUSIONS}
In this work, we have investigated the amplitudes of quantum oscillations of $Al$ and $Be$ point contact resistance in a magnetic field oriented parallel to the contact axis as functions of the potential difference applied to the contact. It was found for the first time that this dependence has a nonmonotonic ascending nature in many cases, and its shape is similar to the EPI spectrum of the point-contact. Such an effect was observed for $Be$ point contacts and for low-resistance $Al$ point contacts. The additional scattering (\emph{in the contact}) of magnetoquantized electrons by phonons generated by accelerated electrons may be a possible reason behind the increase in the amplitude of oscillations with voltage. These processes of scattering of electrons by nonequilibrium phonons increase the contribution to transport of charge carriers through the contact for electrons from extremal Fermi surface cross sections. Hence, in contrast to the conventional suppression of magnetoquantum oscillations due to scattering by impurities in a bulk metal, the oscillations increase with scattering upon an increase in the applied voltage. If, however, scattering by nonequilibrium phonons extends to the contact \emph{banks}, we observe a familiar decrease in the oscillation amplitude (for high-resistance $Al$ contacts). The nonmonotonic $eV$-dependence of the oscillation amplitude $A_1$ (i.e., the existence of a peak near the phonon density of states peak) is probably associated with an energy redistribution of phonons due to phonon-phonon collisions, as well as with the possible energy dependence of the renormalized electron mass in the immediate vicinity of the contact.

The shape of the dependence of the ratio ${{A}_{1}}/{{A}_{2}}$ of the harmonics of oscillation amplitudes on $eV$ can be used to draw conclusions as to the impurity distribution in the quantization region. If the impurity concentration is the same over the entire contact region, the relative intensity of the first harmonic increases with $eV$. If, however, the impurity concentration decreases with increasing distance from the contact center, the opposite effect is observed.

In conclusion, the authors express their sincere thanks to A.N. Omel'yanchuk, E.N. Bogachek, A.A. Zvyagin, and A.V. Dankovskii for stimulating discussions in the course of this research.

This research was partly financed by the State Fund of Fundamental Research, Ukrainian State Committee on Science and Technology, under project No. 2.3/608 (Contact-2).

\end{document}